\documentclass{emulateapj}

\usepackage{graphicx}

\begin{document}

\title{Gamma-ray flares from red giant/jet interactions in AGN}

%\subtitle{II. Star in jets}

\author{Maxim V.~Barkov$^{1,2}$, 
    Felix A.~Aharonian$^{1,3}$
      and Valent\'i Bosch-Ramon$^{1}$}

\affil{$^1$Max-Planck-Institut f\"ur Kernphysik, Saupfercheckweg 1, 69117 Heidelberg, Germany\footnote{bmv@mpi-hd.mpg.de }\\
   $^2$Space Research Institute, 84/32 Profsoyuznaya Street, Moscow, 117997, Russia \\
   $^3$Dublin Institute for Advanced Studies, 31 Fitzwilliam Place, Dublin 2, Ireland}

%\ead{}

\begin{abstract}
{Non-blazar AGN have been recently established as a class of gamma-ray sources. M87, a nearby representative of this class,
show fast TeV variability on timescales of a few days.}
{We suggest a scenario of flare gamma-ray emission in non-blazar AGN based on a
red giant interacting with the jet at the base.}
{We solve the hydrodynamical equations that describe the evolution of the envelope of a red giant blown by the impact of the
jet.}
{If the red giant is at least slightly tidally disrupted by the supermassive black hole, enough stellar material will be blown
by the jet, expanding quickly until a significant part of the jet is shocked. This process can render suitable conditions for
energy dissipation and proton acceleration, which could explain the detected day-scale TeV flares from M87 via proton-proton
collisions. Since the produced radiation would be unbeamed, such an events should be mostly detected from non-blazar AGN.
They may be frequent phenomena, detectable in the GeV-TeV range even up to distances of $\sim 1$~Gpc for the most powerful
jets. The counterparts at lower energies are expected to be not too bright.}
{M87, and nearby non-blazar AGN in general, can be fast variable sources of gamma-rays through red giant/jet interactions.}
\end{abstract}


\maketitle  

%%%%%%%%%%%%%%%%%%%%%%%%%%%%%%%%%%%%%%%%%%%%%%%%%%
\section{Introduction}
\label{intro}

Active galactic nuclei (AGN)  are believed to be powered by an accreting supermassive black hole (SMBH) in the center of a galaxy, 
a significant fraction of AGN show powerful jets, supersonic relativistic flows,  on small (sub parsec) and large 
(multi hundred kpc) scales. \citep[e.g.][]{bbr84}.  These AGN are characterized by nonthermal emission extending from radio to 
high energy gamma-rays. This radiation comes from an accretion disc and from two relativistic jets that are
launched close to the SMBH in two opposite directions. The emission associated to the accretion process can be generated by
thermal plasma in the form of an optically-thick disc under efficient cooling \citep[e.g.,][]{s72,ss73}, or as an
optically-thin corona \citep[e.g.,][]{bkb77,lt79}. The emission from the jets is non thermal and comes from a population of
relativistic particles accelerated for instance in strong shocks, although other scenarios are possible as well \citep[see,
e.g.,][]{recon,na07,rbd07,ra08}. This non-thermal emission is thought to be produced through synchrotron and inverse Compton (IC)
processes \citep[e.g.][]{gmt85}, although hadronic models have been also considered in the past
\citep[e.g.,][]{m93,ah00,mp01,a02}. 

The existence of a stellar clustering in the central regions of AGN, possibly down to very small distances from the central
SMBH \citep[e.g.][]{pens88}, implies that the interaction between a star and the jet should eventually happen.  The gamma-ray
production due to the interaction between an obstacle and an AGN jet has been studied in a number of works. For instance,
\cite{dl97}  suggested the high-energy radiation produced by a beam of relativistic protons impacting with a cloud of the
broad-line region (BLR). The gamma-ray emission from one or many clouds from the BLR interacting with a hydrodynamical
jet recently has bin analyzed by \cite{abr10}. The radiation from the interaction between a massive star and 
an AGN jet was studied in \cite{bp97}. Namely they suggested that the jet interacts with stellar winds of massive stars, 
in their model they assume that the source of gamma-rays is moving with a relativistic speed, 
therefore  the radiation is  Doppler boosted. The main radiation mechanism in this scenario is related to the development 
of the pair cascade in the field of the radiation of the  massive star.  

In this work, we study the interaction of a red giant (RG) star with the base of the jet in AGN and their observable
consequences in gamma rays. We focus here on the case of M87, a nearby non-blazar AGN that presents very high-energy
recurrent activity with variability timescales of few days \citep{ah06,mag,vvhm09,ver10}. In the framework presented here,
the jet impacts the RG envelope, already partially tidaly disrupted by the gravitational field of the central SMBH. The RG
envelope is blown up, forming a cloud of gas accelerated and heated by jet pressure. The jet base is likely strongly
magnetized \citep[e.g.,][]{kbvk07,bk08}. The jet flow affected by the impact with the RG envelope can be a suitable region for
particle acceleration, and a significant fraction of the involved magnetic and kinetic energy of the jet can be transferred
to protons and electrons. Although electrons may not able to reach TeV emitting energies because of the expected large
magnetic fields, protons would not suffer from this constraint. These protons could reach the star blown material, and
optically-thick proton-proton ($pp$) interactions could lead to significant gamma-ray production in the early stages of the
cloud expansion. Unlike in \cite{bp97}, we deal with solar-mass-type 
stars instead of the more rare high-mass stars, study
the RG atmosphere-jet interaction, and follow the hydrodynamical evolution of the cloud. Finally, we do not introduce any
beaming factor to the radiation, since in our scenario most of the emission is produced when the cloud has not been
significantly accelerated, Doppler boosting being therefore negligible.

%%%%%%%%%%%%%%%%%%%%%%%%%%%%%%%%%%%%%%%%%%%%%%%%%%
\section{The model}
\label{model}

{  Main sequence stars are too compact to be significantly affected by tidal forces from the SMBH, unlike RGs, whose external 
layers are far less gravitationally bounded to the stellar core. Therefore,
in the vicinity of a SMBH, the external layers of an RG will suffer significant tidal disruption
\citep[see][]{knp93a,knp93b,dfkn97,alp00,icn03,lkp09}, which can unbound from the stellar core 
a cloud with significant mass $\gtrsim 10^{30}$~g. 
Therefore, if an RG penetrates into the innermost region of the jet, 
the RG envelope can be already weakly gravitationally attached to the star due to tidal disruption. In this situation, the
external layers of the star can be lost due to jet ablation, which is unlikely in the case of undisrupted RGs (except for 
very powerful jets). 

The tidal forces are important when the distance between
the SMBH and the star is similar to or smaller than the tidal distance ($z_{\rm T}$) for a given RG radius ($R_{\rm RG}$) and
mass ($M_{\rm RG}$)}, where:
\begin{equation}
z_{\rm T}=R_{\rm RG}\left(\frac{M_{\rm BH}}{M_{\rm RG}}\right)^{1/3},
\label{rt}
\end{equation}
and $M_{\rm BH}$ is the mass of the SMBH. Therefore, for a given RG-jet interaction distance to the SMBH
$z$, the RG can lose the atmosphere layers beyond $R_{\rm *T}$. For the case of M87, with $M_{\rm
BH}=(6.4\pm0.5)\times 10^9\,M_{\odot}$ \citep{gt09}, one obtains:
\begin{equation}
R_{\rm RG}^{\rm T}=z\,\left(\frac{M_{\rm RG}}{M_{\rm BH}}\right)^{1/3}\approx 
76\,M_{\rm RG\odot}^{1/3}\,R_{\odot}\approx 5.3\times 10^{12}\,M_{\rm RG\odot}^{1/3}\,\mbox{cm}\,,
\label{rst}
\end{equation}
% I have remove _{16}, it is not needed here
where $M_{\rm RG\odot}\equiv M_{\rm RG}/1\,M_\odot$. Since a solar-mass RG, the most common one, can have up to few hundreds of
$R_\odot$, a significant fraction of the star envelope can be carried away by the jet flow up to $z\la 10^{17}$~cm. Note that
evidence for the presence of a radio jet has been found from M87 within a distance of $\sim 10^{17}$~cm from the SMBH
\citep{jun99}.

The M87 TeV lightcurve obtained by \cite{ah06} shows several peaks, and each of these peaks in our model correspond to different RG-jet
events. Note however that some nearby peaks may correspond to a complex disruption process, motivated for instance by a very
disrupted and massive envelope, or by jet inhomogeneities. Also, it cannot be discarded that a cluster of several RGs
could also enter the jet.

The time needed by the RG to cross the jet cannot be shorter than the typical M87 event duration, $t_{\rm e}\sim  2\times
10^5$~s. It cannot be longer either, since then the event duration would also be longer if there is
available RG matter for removal, as expected at $z\la z_{\rm T}$. 
Therefore, $t_{\rm jc}=t_{\rm e}$, and the
interaction height can be derived taking the velocity of the RG orbiting the SMBH as the Keplerian velocity:
\begin{equation}
z_{\rm jc}=\left[ G\,M_{\rm BH}\left(\frac{t_{\rm jc}}{2\theta} \right)^2 \right]^{1/3}
\approx  10^{16} \theta_{-1}^{-2/3}\mbox{cm},
\label{rjct}
\end{equation} 
where $\theta_{-1}=\theta/0.1$ is the jet semi-opening angle in radians.  An important parameter is the power of the jet  
$L_{\rm j}\approx 1-5\times 10^{44}$ erg s$^{-1}$ \citep{oek00}, which
we fix to $L_{\rm j}\approx 2\times 10^{44}$ erg s$^{-1}$. From $L_{\rm j}$ and the jet width, $z_{\rm
jc} \theta$, we can derive the jet energy flux at the interaction height:
\begin{equation}
F_{\rm j}=\frac{L_{\rm j}}{\pi z_{\rm jc}^2\theta^2} \approx 10^{14} \mbox{erg~cm}^{-2} \mbox{ s}^{-1}.
\label{lir}
\end{equation}

\begin{figure} 
\includegraphics[width=80mm,angle=-0]{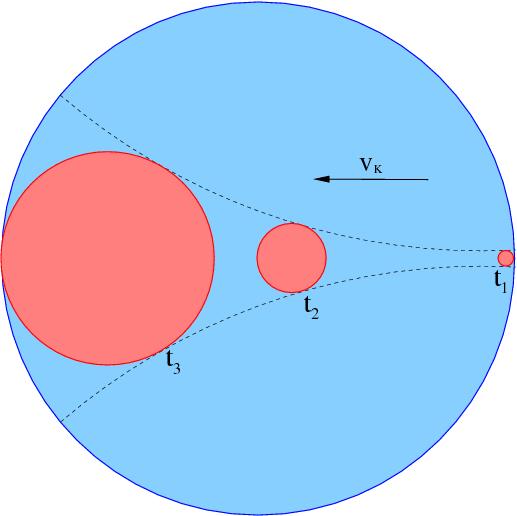}
\caption{Sketch of the 
evolution within the jet of the cloud formed by the disrupted envelope of the RG. The plane of the image would be the jet 
section.}
\label{clouds}
\end{figure}

There are two regimes for the RG tidal disruption: under strong tidal interaction ($R_{\rm RG}>R_{\rm *T}$), the RG envelope
suffers an elongation along the direction of motion of the star \citep{knp93b}; under weak tidal interaction ($R_{\rm RG}\sim
R_{\rm *T}$), the envelope is still roughly spherical \citep{knp93a}. In both situations, the outer layers of the star will
be swept away by the jet, forming a cloud that will quickly heat up and expand. We study the time evolution of the cloud
adopting a very simplified hydrodynamical model for the cloud expansion. 
The heating of the cloud is caused by the propagation
of shock waves, which are formed by the pressure exerted by the jet from below.  
Therefore, the cloud pressure is taken similar
to the jet pressure (regardless it is of kinetic or magnetic nature): 
\begin{equation}
p_{\rm j}=\frac{F_{\rm j}}{c}\approx p_{\rm c}\approx(\hat\gamma-1)e_{\rm c}\,,
\end{equation}
where $c$ is the speed of light, 
and $\hat\gamma$ the adiabatic index ($\hat\gamma=4/3$).
The cloud expands at
its sound speed ($c_{\rm s}$), since the lateral and top external pressures are much smaller than the jet bottom one. 
{  When the cloud has significantly expanded, its pressure becomes smaller than the jet pressure from below. 
At that point new shocks
%develop in the contact discontinuity 
leading to further cloud heating. We
illustrate in Fig.~\ref{clouds}, for the simplest case of weak disruption, how the spherical cloud} evolves under the effect of
jet pressure as seen in the plane perpendicular to the jet axis.
% we look from Z direction
% it is not the evolution in Z direction
% in the $z$-direction.

 %fffffffffffffffffffffffffffffffffffffffffffffffffffffffffffffffff
 %fffffffffffffffffffffffffffffffffffffffffffffffffffffffffffffffff

 %fffffffffffffffffffffffffffffffffffffffffffffffffffffffffffffffff
\begin{figure} 
\includegraphics[width=80mm,angle=-0]{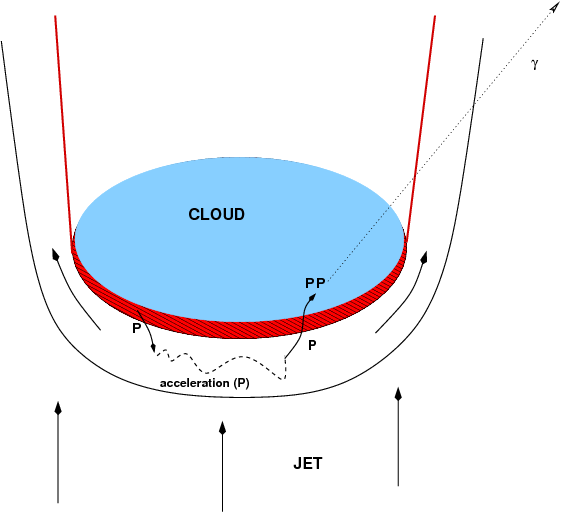}
\caption{Sketch of the proton acceleration and gamma-ray production processes.
The plane of the image would be normal to the jet section.}
\label{prot_acc}
\end{figure}
 %fffffffffffffffffffffffffffffffffffffffffffffffffffffffffffffffff

\subsection{Weak tidal interaction (spherical case)}
  
The system of equations that characterizes the weak tidal interaction case can be written as follows:
\begin{equation}
E_{\rm c}=e_{\rm c} \frac{4\pi r_{\rm c}^3}{3}= \frac{4 \pi F_{\rm j} r_{\rm c}^3}{3(\hat\gamma-1)c}
\label{scl_e}
\end{equation}
\begin{equation}
\frac{dr_{\rm c}}{dt}=c_{\rm s}=\left(\frac{\hat\gamma(\hat\gamma-1)E_{\rm c}}{M_{\rm c}}\right)^{1/2}
\label{scl_r}
\end{equation}
\begin{equation}
\frac{d^2z_{\rm c}}{dt^2}=\frac{\pi F_{\rm j} r_{\rm c}^2}{cM_{\rm c}}\,,
\label{scl_z}
\end{equation}
where $r_{\rm c}$ and $M_{\rm c}$ are the cloud radius and mass, respectively. 

\begin{figure}
\includegraphics[width=84mm,angle=-0]{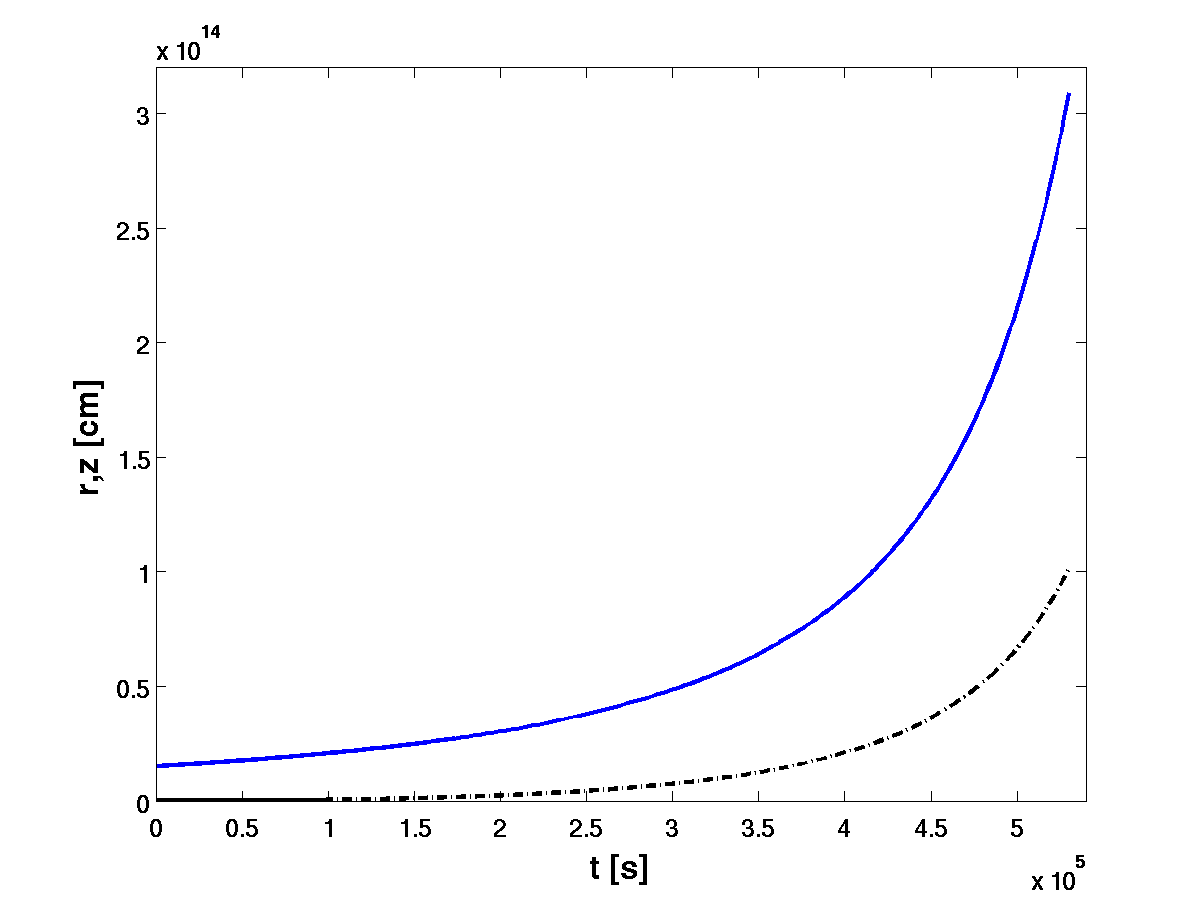}
\caption{Evolution of $r_{\rm c}$ (solid line) and $z-z_{\rm jc}$ (dot-dashed lime) with time in the weak tidal interaction 
case. The parameter values are
characteristic of  M87: $L_{\rm j}=2\times 10^{44}$~erg~s$^{-1}$, 
$M_{\rm BH}=6.4\times 10^9\,M_\odot$, $\theta_{-1}=0.5$, $M_{\rm RG}=1\,M_\odot$, $z_{\rm jc}\approx 2.5 \times 10^{16}$~cm, and
$M_{\rm c}\approx 1.3\times 10^{28}$~gr.}
\label{rzs}
\end{figure}

The solutions to Eqs.~(\ref{scl_e}-\ref{scl_z}) are:
\begin{equation}
r_{\rm c}(t)=\frac{r_{\rm c0}}{(1-t/t_{\rm ce})^2 }
\label{rcl_j}
\end{equation} 
\begin{equation}
v_r(t)=\frac{2r_{\rm c0}}{t_{\rm ce}(1-t/t_{\rm ce})^3}\,,
\label{vrcl_j}
\end{equation} 
where $r_{\rm c0}$, assumed to be similar to $R_{\rm *T}$, 
is the initial cloud radius, and $t_{\rm ce}$ the cloud characteristic 
expansion time:
\begin{equation}
t_{\rm ce} = \left(\frac{3 c M_{\rm c}}{\pi\hat\gamma F_{\rm j}r_{\rm c0}}\right)^{1/2}
\approx 5\times 10^5 \left(M_{\rm c28}/F_{j,14}r_{\rm c0,13}\right)^{1/2}\,{\rm s}\,,
\label{Acl}
\end{equation}
where $M_{\rm c28}=M_{\rm c}/10^{28}\,{\rm g}$.
Neglecting the initial cloud velocity in the $z$-direction, we obtain:
\begin{equation}
z(t)-z_{\rm jc}=\frac{r_{\rm c0}}{2\hat\gamma}\left(\frac{t}{t_{\rm ce}}\right)^{2} 
\frac{3/2-t/t_{\rm ce}}{(1-t/t_{\rm ce})^2}
\label{rcl_z}
\end{equation}
\begin{equation}
v_z(t)=\frac{r_{\rm c0}}{\hat\gamma t_{\rm ce}}
\frac{(t/t_{\rm ce})^2((t/t_{\rm ce})^2-3t/t_{\rm ce}+3)}{(1-t/t_{\rm ce})^3}
\label{rcl_vz}
\end{equation}
where $z_{\rm jc}$ is the RG-jet penetration height. The evolution of the cloud radius is presented in Fig.~\ref{rzs}. The
adopted parameter values are: $L_{\rm j}=2\times 10^{44}$~erg~s$^{-1}$,  $M_{\rm BH}=6.4\times 10^9\,M_\odot$,
$\theta_{-1}=0.5$, $M_{\rm RG}=1\,M_\odot$, $z_{\rm jc}\approx 2.5 \times 10^{16}$~cm, $M_{\rm c}\approx 1.3\times
10^{28}$~gr. Note that for times $t < t_{\rm ce}$, $z-z_{\rm jc}\ll r_{\rm c}<\theta\,z_{\rm jc}$ and $v_z\ll c_{\rm s}\ll
c$.

%fffffffffffffffffffffffffffffffffffffffffffffffffffffffffffffffff
%fffffffffffffffffffffffffffffffffffffffffffffffffffffffffffffffff 

Making the cloud and jet pressures comparable, the energy transfer 
can be overestimated
beyond a certain radius ($r_{\rm ct}$) and time
($t_{\rm t}$) during the cloud evolution, in which energy balance is to be fulfilled: 
\begin{equation}
\frac{dE_{\rm c}}{dt}\le\pi r_{\rm c}^2 F_{\rm j}\,.
\label{eb}
\end{equation}
Using Eqs.~(\ref{rcl_j}) and (\ref{vrcl_j}), (\ref{eb}) permits to derive $t_{\rm t}$:
\begin{equation}
t_{\rm t}=t_{\rm ce}\left(1-\frac{8}{\hat\gamma-1}\frac{r_{\rm c0}}{t_{\rm ce} c}\right)\,.
\label{tb}
\end{equation}
Substituting Eq.~(\ref{tb}) into Eq.~(\ref{rcl_j}), we obtain $r_{\rm ct}$:
\begin{equation}
r_{\rm ct}=\frac{r_{\rm c0}}{\left(\frac{8}{\hat\gamma-1}\frac{r_{\rm c0}}{t_{\rm ce} c}\right)^{2/3} }
\approx 1.5\times 10^{14} M_{\rm c,28}^{1/3}F_{j,14}^{-1/3} \mbox{ cm}\,.
\label{rb}
\end{equation}
Thus, if $r_{\rm c}<r_{\rm ct}$, the solutions presented in Eqs.~(\ref{scl_e})-(\ref{scl_z}) are valid. After $t_{\rm t}$, 
$dE_{\rm c}/dt\sim \pi r_{\rm c}^2 F_{\rm j}$ yields a slower increase of $r_{\rm c}$ with $t$, although this should be
still a fast exponential growth. We consider $t<t_{\rm t}$ since we focus here on the case when the cloud
is optically thick to $pp$ collisions (see below), and at $t>t_{\rm t}$ the cloud density is already too low. 

\subsubsection{Strong tidal interaction (elongated case)}

{  In the case of strong tidal interaction the RG atmosphere is stretched in the direction of motion of the star, and the 
expansion will be now cylindric}. In such a
case, $R_{\rm RG}^{\rm T}=r_{\rm c0}$ ($r_{\rm c}$ is the cloud cylindrical radius) can be significantly smaller than the
length of the disrupted atmosphere, $l_{\rm c}$ \citep{alp00}.  The system of equations describing this case can be written
as:
\begin{equation}
E_{\rm c}=\frac{\pi l_{\rm c} r_{\rm c}^2 F_{\rm j}}{(\hat\gamma-1)c}
\label{ccl_e}
\end{equation}
\begin{equation}
\frac{dr_{\rm c}}{dt}=c_{\rm s}=\left(\frac{\hat\gamma(\hat\gamma-1)E_{\rm c}}{M_{\rm c}}\right)^{1/2}
\label{ccl_r}
\end{equation}
\begin{equation}
\frac{d^2z_{\rm c}}{dt^2}=\frac{2 l_{\rm c} F_{\rm j} r_{\rm c}}{c M_{\rm c}}\,.
\label{ccl_z}
\end{equation}

Substituting Eq.~(\ref{ccl_e}) into Eq.~(\ref{ccl_r}), we obtain:
\begin{equation}
r_{\rm c}(t)=r_{\rm c0} e^{t/t_{\rm ce}}.
\label{rccl_j}
\end{equation} 
As in the weak case, $r_{\rm c0}$ and $t_{\rm ce}$ are the initial radius and the expansion time of the cloud, where:
\begin{equation}
t_{\rm ce}=\left(\frac{cM_{\rm c}}{\pi \hat\gamma F_{\rm j} l_{\rm c}} \right)^{1/2}
=1 M_{\rm c,28}^{1/2} F_{\rm j,14}^{-1/2} l_{\rm c,14}^{-1/2}\,\mbox{day}\,,
\label{tde}
\end{equation} 
with $l_{\rm c,14}=(l_{\rm c}/10^{14}\,{\rm cm})$ and $F_{\rm j,14}=(F_{\rm j}/10^{14}\,{\rm erg~cm}^{-2}{\rm s}^{-1})$

If we neglect the initial velocity of the cloud in the $z$-direction, the distance covered by the cloud is as follows:
\begin{equation}
z(t)-z_{\rm jc}=\frac{2 F_{\rm j} l_{\rm c} r_{\rm c0} t_{\rm ce}}{c M_{\rm c}} 
\left( t_{\rm ce} e^{t/t_{\rm ce}}-t_{\rm ce}-t\right)\,,
\label{rccl_z}
\end{equation}
with a velocity
\begin{equation}
v_z(t)=\frac{2 F_{\rm j} l_{\rm c} r_{\rm c0} t_{\rm ce}}{c M_{\rm c}} \left( e^{t/t_{\rm ce}}-1\right)\,.
\label{rccl_vz}
\end{equation}

As in the weak case, after substantial expansion equalizing cloud and jet pressures overestimates the energy transfer from
the jet to the elongated cloud, and the relation $dE_{\rm c}/dt\sim \pi r_{\rm c}^2 F_{\rm j}$ should be used. This phase is
characterized by a slower, but still quite fast, power-law-like expansion rate.

\section{Radiation}

{  Particles could be accelerated in the shocked jet region below the cloud.}
As noted in Sect.~\ref{intro}, the jet is probably magnetically dominated at $z\la z_{\rm T}$. 
Therefore, one can estimate the magnetic field in the jet as follows:
\begin{equation}
B_{\rm j}\approx \sqrt{\frac{8 L_{\rm j}}{c z^2 \theta^{2}}}
\approx 200\,L_{\rm j,44}^{1/2}z_{16}^{-1}\theta_{-1}^{-1}\,\mbox{G}\,,
\label{bj}
\end{equation}
where $L_{\rm j,44}=L_{\rm j}/10^{44}\,{\rm erg~s}^{-1}$. {  The expected magnetic field in the shocked jet region should 
be also strong, probable of a similar strength to $B_{\rm j}$.}
Under such a magnetic field, one can estimate the acceleration timescale:
\begin{equation}
t_{\rm acc}=\frac{E}{\dot{E}_{\rm acc}}\sim \frac{\xi\,E}{q\,B_{\rm j}\,c\,}\approx
0.1\,\xi\,E_2\,B_{\rm j,2}^{-1}\,{\rm s}\,,
\label{acc1}
\end{equation}
where $\xi$ is the acceleration efficiency parameter, $q$ is the particle charge, 
$E_2=E/10^2~{\rm TeV}$, and $B_{\rm j,2}=B_{\rm j}/10^2\,{\rm G}$,
the maximum energy of protons and electrons are 
\begin{equation}
E_{\rm p~max}\approx \sqrt{\frac{3}{2\xi}}\,q\,B_{\rm j}\,r_{\rm c}\approx 
10^7\,B_{\rm j,2}\,r_{\rm c,14}\xi^{-1/2}\,\,{\rm TeV}\,
\label{acc2}
\end{equation}
and
\begin{equation}
E_{\rm e~max}\approx \sqrt{\frac{q\,c}{\xi\,a_{\rm s}\,B_{\rm j}}}\approx 
10\,B_{\rm j,2}^{-1/2}\,\xi^{-1/2}\,{\rm TeV}\,,
\label{acc3}
\end{equation}
respectively, where $a_{\rm s}=1.6\times 10^{-3}$. Equation~(\ref{acc2}) is obtained from limiting the proton acceleration by
Bohm diffusion escape from the interaction region, of size $r_{\rm c,14}=(r_{\rm c}/10^{14}\,{\rm cm})$, and Eq.~(\ref{acc3}) 
is
obtained from limiting the electron acceleration through synchrotron cooling. 
Even taking a high $\xi\sim 10$ 
(for mildly relativistic shocks, as those of supernova explosions, $\xi\sim 10^4$), electron
energies will be too low to explain the HESS spectrum of M87 up to energies of few 10~TeV \citep{ah06}, whereas protons
may be accelerated up to ultra-high energies. 
In addition, the expected $B_{\rm jet}$-values could easily suppress
any IC component. 
{  We note that even for diffusion faster than Bohm, or under bigger $\xi$-values, protons could still reach
enough energy to explain observations.}
On the other hand, the cloud density can be high, making of $pp$ interactions the best candidate for
gamma-ray production in the RG-jet scenario, the characteristic cooling time for $pp$ collisions being:
\begin{equation}
t_{pp}\approx \frac{10^{15}}{n_{\rm c}}=10^5\,n_{\rm c,10}^{-1}\,{\rm s}\,,
\label{pp}
\end{equation}
where $n_{\rm c,10}=n_{\rm c}/10^{10}\,{\rm cm}^{-3}$ is the cloud density. We note that the high cloud density should not
affect significantly the proton acceleration, which would occur in the far less dense jet shocked region. {  Nevertheless,
protons should penetrate in the acceleration process and, in the Blanford-Znajek scenario of jet formation \citep{BZ77,BIP92}
the jet is probably formed only by pairs at $z_{\rm jc}$. Therefore, 
some cloud material should penetrate into the shocked jet medium, which can
occur through Rayleigh-Taylor instabilities \citep{Chandra61,Imsh72}. We present in Fig.~\ref{prot_acc} a sketch of the mixing, 
proton acceleration
and gamma-ray production processes. We do not specify here the physics of particle acceleration, 
although this could take place by one or a combination of different mechanisms: magnetic reconnection right after the
shock in the jet, shear acceleration due to the strong velocity gradients close to the contact discontinuity, or some sort 
of stochastic acceleration due to magnetic turbulence downstream the jet shock.} 
Regarding other proton radiation mechanisms, proton synchrotron will not be efficient in
our case, with $t_{\rm p~sync}\approx 5\times 10^{10}\,B_{\rm j,2}^{-2}\,{\rm s}\gg t_{pp}$. Also photomeson production can
be also neglected, since $t_{\rm p\gamma}\sim 5\times 10^6\,L_{\rm 41}\,r_{\rm c,14}^{-1}\,\epsilon_{\rm keV}^{-1}{\rm s}\gg
t_{pp}$, where $L_{\rm 41}=(L/10^{41}\,{\rm erg~s}^{-1})$ and $\epsilon_{\rm kev}=(\epsilon/1~{\rm keV})$ would be the
luminosity produced in the region and the ambient photon energy (e.g. thermal X-rays; see below), respectively. photomeson
production with keV ambient photons would require protons with energies above $\sim 100$~TeV

Hereafter, we will treat the generation of protons with energies $>100$~GeV phenomenologically, assuming that a fraction
$\eta$ of the total eclipsed jet luminosity is converted to relativistic protons:  $L_{\rm p}=\eta\pi\,r_{\rm c}^2\,F_{\rm
j}$. We also assume that these protons can effectively reach the cloud, where they suffer $pp$ interactions that lead to
$\pi^0$-meson production, {  although some of them could escape surrounding the cloud.} Once in the cloud, protons can be effectively trapped during the RG-jet interaction time for
magnetic fields as low as few G, since:
\begin{equation}
B_{\rm c}=\frac{t\,E\,c}{3\,q\,r_{\rm c}^2}\approx 0.03\,t_{5}\,E_2\,r_{\rm c,14}^{-2}\,{\rm G}\,,
\label{bb}
\end{equation}
where $t_{5}=t/10^5\,{\rm s}$ is the time inside the cloud. {  Note that Eq.~(\ref{bb}) has been derived assuming Bohm diffusion, but
faster diffusion regimes would still allow to keep protons trapped in the cloud.}

The typical fraction of the proton energy transferred 
per collision to the leading gamma rays is $E_{\gamma}=0.17 E_p$ \citep{kab06} in the optically thin case, and around twice
that value for optically thick media neglecting gamma-rays from other secondary particles. Therefore, we can characterize 
the proton-gamma ray energy transfer by
\begin{equation}
\chi\equiv E_{\gamma}/E_p = 0.17\,[2-\exp(-t/t_{pp})]\,.
\label{enef}
\end{equation}

Two phases of the cloud expansion can be distinguished: the radiatively efficient regime, i.e. with $\chi\approx 0.34$ or
$t>t_{pp}$, and the radiatively inefficient regime, with $\chi=0.17$ or $t<t_{pp}$. Thus, from the simplifications above,
the gamma-ray luminosity in the $pp$ optically-thick case can be written as:
\begin{equation}
L_{\gamma}\approx 0.34\,\eta\,\pi\,r_{\rm c}^2\,F_{\rm j}\,,
\label{egc}
\end{equation} 
where is seen that $L_\gamma\propto r_{\rm c}^2$.
In the $pp$ optically thin case, only a fraction $t/t_{pp}$ of $L_{p}$ is lost through $pp$ collisions, and
$L_{\gamma}\propto r_{\rm c}^{-1}$. 
The general expression for the gamma-ray luminosity during a RG-jet
interaction event becomes:
\begin{equation}
L_{\gamma}\approx \pi\eta\chi\,r_{\rm c}^2\,F_{\rm j}\,(1-\exp(-t/t_{pp}))\,. 
\label{egcm}
\end{equation} 
Given the fast expansion of the cloud, either in the spherical or the elongated case,
one can expect a sharp spike in the light curve.

Secondary electrons and positrons ($e^\pm$), injected by $pp$ collisions with an energy rate $\sim L_\gamma$, could emit most
of their energy through synchrotron radiation. Given the moderate energy budget, the radio, optical and X-ray fluxes would be
below the observed values in the region of interest. However, at later times, conditions may change becoming more suitable
for radio emission. Thus, it cannot be discarded that RG-jet interactions could eventually have a low-energy faint
counterpart produced by secondary or thermal $e^\pm$, { or even by a primary population of accelerated electrons.}

\begin{figure}
\includegraphics[width=84mm,angle=-0]{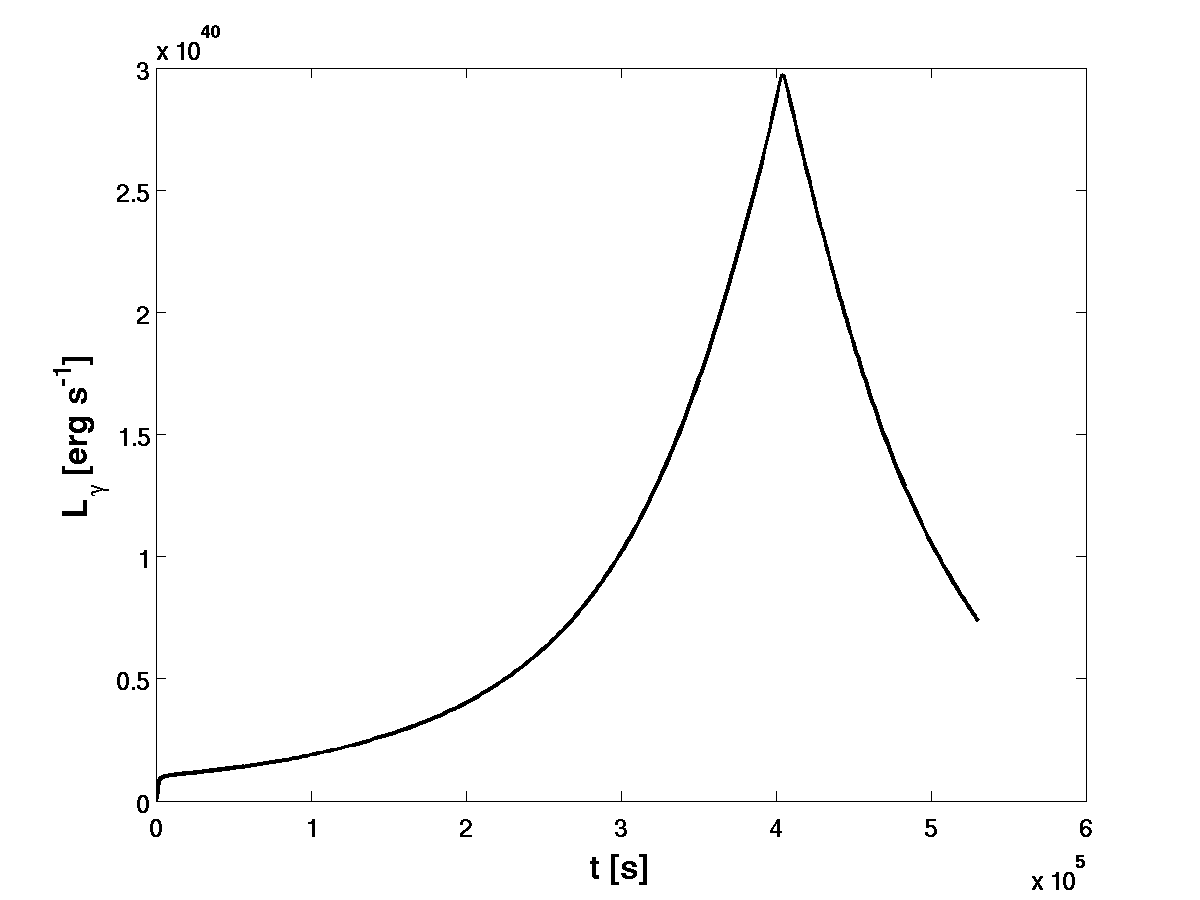}
\caption{Gamma-ray $pp$ lightcurve for the weak tidal disruption case. The same parameter values as in Fig.~\ref{rzs} have 
been adopted.}
\label{degdts}
\end{figure}

The gamma-ray $pp$ light curve for M87 is presented in Fig.~\ref{degdts}, in which the maximum is reached at $t_{\rm
peak}\approx 4\times 10^5$~s, with a width of $\sim 1-2$~days. A value for $\eta$ of 0.1 has been adopted. We recall that our
approach is valid for $t<t_{\rm t}\approx 4.7\times 10^5$~s (see Sect.~\ref{model}), hence the adopted cloud evolution model
describes the gamma-ray peak properly. We show the gamma-ray lightcurve for the weak tidal disruption case as the most
conservative scenario. In the strong tidal disruption case, the lightcurve would be similar but the gamma-ray maximum would
be even higher.  

 %fffffffffffffffffffffffffffffffffffffffffffffffffffffffffffffffff
%fffffffffffffffffffffffffffffffffffffffffffffffffffffffffffffffff  

\subsection{Optimal radiation case}

Around the gamma-ray maximum, at $t\sim t_{\rm peak}$, and because of $t_{\rm peak}\sim t_{pp}$, one can measure the cloud
density and radius $n_{\rm cp}$ and $r_{\rm cp}$. From
\begin{equation}
n_{\rm cp}=\frac{3\,M_{\rm c}}{4\pi\,m_p\,r_{\rm cp}^3}\,;
\label{ntv}
\end{equation}
plus the variability time:
\begin{equation}
t_{\rm v}\approx t_{pp} \approx \frac{10^{15}}{n_{\rm cp}} 
\label{tvtpp}
\end{equation}
one can determine 
\begin{equation}
r_{\rm cp}=\left(\frac{3M_{\rm c}\,t_{\rm v}}{4\pi\,10^{15}\,m_p}\right)^{1/3}\,.
\label{rc}
\end{equation}
%However, be possible sensitivity limitations of present Cherenkov telescopes may not allow the observation of the earlier expansion times
%of the cloud, preventing a proper determination of $t_{\rm peak}$. Therefore, 

We can characterize the variability time $t_{\rm v}$, which
would correspond to the characteristic duration of the gamma-ray peak, as follows:
\begin{equation}
t_{\rm v}=2\frac{L_\gamma}{dL_\gamma/dt}=\frac{t_{\rm ce}(1-t_{\rm peak}/t_{\rm ce})}{2}\,.
\label{tv}
\end{equation} 
The condition $r=r_{\rm cp}$, and Eqs.~(\ref{rcl_j}), (\ref{rc}) and  (\ref{tv}), allow the derivation of 
the following expression:
\begin{equation}
\left(\frac{48M_{\rm c}}{\pi\,10^{15}\,m_p}\right)^{1/3}t_{\rm v}^{7/3}=t_{\rm ce}^2 r_{\rm c0}\,.
\label{exp}
\end{equation} 
Then, from Eq.~(\ref{exp}), and assuming $t_{\rm ce}=t_{\rm jc}$, $(1-t_{\rm peak}/t_{\rm ce})\ll 1$ and $r_{\rm c0}=R_{\rm
RG}^{\rm T}$, one obtains:
\begin{equation}
t_{\rm ce}= \frac{L_{\rm j}^{3/20} \hat\gamma^{3/20} \theta^{1/10} t_{\rm v}^{21/20} }{
50\times 10^{1/4} \pi^{3/20} c^{3/20} G^{1/5} M_{\rm BH}^{1/10} m_p^{3/20} M_{\rm RG}^{1/10}}
\label{tcer}
\end{equation}
\begin{equation}
M_{\rm c}= \left(\frac{L_{\rm j}^{6} \hat\gamma^{6} M_{\rm RG} t_{\rm v}^{7} }{
 750^5\pi c^{6} G^{3} M_{\rm BH}^{4} m_p^{} \theta^6}\right)^{1/5}
\label{mb}
\end{equation}
\begin{equation}
z_{\rm jc}=
\label{rjc}\left(\frac{
M_{\rm BH}^4 G^3 L_{\rm j}^{3/2}\hat\gamma^{3/2} t_{\rm v}^{21/2}
}{
3.16\times 10^{22}\pi^{3/2} m_p^{3/2} c^{3/2} M_{\rm RG} \theta^9
}\right)^{1/15}.
\end{equation}
Finally, substituting (\ref{rjct}), (\ref{lir}), (\ref{rc}) and (\ref{mb}) into (\ref{egc}), the expression for the gamma-ray
luminosity around the lightcurve peak can be derived:
\begin{eqnarray}
L_{\gamma}&=&\frac{\eta\chi}{10^9 \pi^{3/5}} \left(\frac{M_{\rm RG}}{M_{\rm BH}^4}\right)^{4/15} 
\left( \frac{\hat\gamma^3 L_{\rm j}^{8}t_{\rm v}}{c^3G^4m_p^3\theta^8} \right)^{1/5}
\nonumber\\{} & \approx & 
8\times 10^{40}\eta_{-1} L_{j,44}^{8/5}t_{v,5}^{1/5} M_{\rm RG}^{4/15} M_{\rm BH,9}^{-16/15} \theta_{-1}^{-8/5}\,
\mbox{erg s}^{-1},
\label{egcn}
\end{eqnarray}
where $M_{\rm BH,9}=(M_{\rm BH}/10^9\,M_\odot)$, $\eta_{-1}=\eta/0.1$ and $t_{\rm v,5}=(t_{\rm v}/10^5\,{\rm s})$.
Adopting typical parameter values for M87, $L_{\rm j}=2\times 10^{44}$~erg~s$^{-1}$, $M_{\rm BH}=6.4\times 10^9\,M_\odot$,
$\theta_{-1}=1$, $t_{\rm v}=2\times 10^5$~days, $M_{\rm RG}=1\,M_\odot$, $z_{\rm jc}\approx 3.6 \times 10^{16}$~cm, 
$M_{\rm c}\approx 1.4\times 10^{28}$~gr, and $\eta_{-1}=1$, 
one gets $L_\gamma\approx 4\times 10^{40}\,\mbox{erg s}^{-1}$, in good agreement  with
observations \citep{ah06,mag,vvhm09,ver10}. We remark that the $t_{\rm v}$-value would be in agreement with
the observed event durations.

\subsection{Thermal radiation of the cloud and self-$\gamma\gamma$ absorption}

In the case of M87, near the peak of the very high-energy (VHE) radiation, i.e. 
$n_{\rm cp}\approx 10^{10} $~cm$^{-3}$ and $r_{\rm cp}\approx 10^{14}$~cm, 
the cloud is optically thin to the radiation produced by its own shocked plasma:
\begin{equation}
\tau_{e\gamma}=r_{\rm cp}\,n_{\rm cp}\,\sigma_{\rm T}\approx 0.6<1, 
\label{taut}
\end{equation}
where $\sigma_{\rm T}=6.65\times10^{-25}$ cm$^{-2}$ is the Thompson cross section. 
At the temperature of the shocked cloud, $T_{\rm c}\sim 10^{10}$~K, the timescales for Coulombian thermalization through
$p-p$ and $e-e$ scattering are $t_{e-e}\approx t_{p-p}\approx  T_8^{3/2} n_{\rm c10}^{-1}\approx 1000$~s ($t_{ep}\approx
10^3\,T_8^{3/2}\,n_{\rm c10}^{-1}\approx 10^6$~s for $p-e$ scattering). Therefore, the shocked cloud is thermalized.

The main channel of thermal radiation is free-free emission, with a photon mean energy $<\epsilon>\,\sim k T_{\rm c}\sim 1$~MeV
and total luminosity \citep{blp71,kp79}:
\begin{eqnarray}
L_{\rm X}&=& 2.1\times 10^{-27} T^{1/2} n_{\rm c}^2 V_{\rm cp} \approx 10^{41}\,\mbox{erg~s}^{-1},
\label{lx}
\end{eqnarray} 
where $V_{\rm cp}=4\pi r_{\rm cp}^3/3$. The concentration of thermal photons can be estimated as
\begin{eqnarray}
n_{\rm X}&=& \frac{L_{\rm X}}{4\,\pi\,c\,k\,T_{\rm c}\,r_{\rm cp}^2}\approx 2 \times 10^7\,\mbox{cm}^{-3},
\label{nx}
\end{eqnarray}  
yielding an optical depth for photon-photon absorption at the energy of the strongest attenuation ($\sim m_e^2\,c^4/<\epsilon>$):
$\tau_{\gamma\gamma}\sim 0.2\,n_{\rm X}\,r_{\rm cp}\,\sigma_{\rm T}\approx 10^{-3}\ll 1$ \citep{ah04}, being much smaller at
1~TeV. The free-free radiation should not show any thermal lines, presenting a very hard non-thermal X-ray spectrum.

For very powerful jets the condition presented in Eq.~(\ref{taut}) is not fulfilled, the shocked plasma is radiation
dominated and cooler, and Eqs.~(\ref{lx}) and (\ref{nx}) do not apply. The cloud is then optically thick, with the radiation
being a black body with mean energy of photons $<\epsilon>\approx 3 k\,T_{\rm b}\approx 10 L_{\rm j,44}^{1/5}\,M_{\rm
BH,9}^{-2/15}\,t_{\rm v,5}^{-21/60}$~eV. The optical depth for gamma-rays is $\tau_{\gamma\gamma}\approx 10^4 L_{j,44}^{3/5}
M_{\rm BH,9}^{1/15} t_{\rm v,5}^{-26/15}$, with the radiation being suppressed for energies $E_{\rm th}\ga
m_e^2\,c^4/3\,k\,T_{\rm b}\sim 50$~GeV, where $m_e$ is the electron mass. Photon-photon absorption creates pairs with
energies $\ga E_{\rm th}$ that cool down through synchrotron emission with spectral energy distribution $\epsilon
F_{\epsilon}\propto \epsilon^{1/2}$ below 10~keV, with the higher energy part of the spectrum softer, reaching MeV-GeV
energies. Under $\tau_{\gamma\gamma}>1$ and reasonable magnetic fields, the synchrotron luminosity will be similar to the
absorbed gamma-ray luminosity.

The X-ray flare detected from M87 almost simultaneously with the VHE flare \citep[see, e.g.,][]{vvhm09} may have been also
produced at the RG-jet interaction. This X-ray flare could be of synchrotron nature, with possible contributions from a
primary electron component, secondary $e^\pm$ from $pp$ and photon-photon interactions, and thermal free-free radiation.
Regardless the origin, the observed X-ray emission could have a counterpart at lower energies. If it came from the region of
gamma-ray production, the spectrum should be quite hard to avoid too many optical photons that would lead otherwise to
significant TeV photon absorption. Optical observations simultaneous with a gamma-ray flare could clarify this point. 

%$\tau_{e\gamma}\approx 0.5\,M_b^{1/3}\,t_v^{-2/3}$ 

%\begin{eqnarray}
%\varepsilon_{max} \approx \frac{(m_ec^2)^2}{<hv>} \approx 50 L_{j,44}^{-1/5} M_{\rm BH,9}^{2/15} t_{\rm v,5}^{21/60} \mbox{ GeV}.
%\label{egg}
%\end{eqnarray} 
%The TeV photons will prodise pairs, which will radiate in 1-1000 MeV rage with spectra $F_v\propto v^{-1/2}$.
% Please check this...

%ssssssssssssssssssss
\section{Discussion and conclusions}
%ssssssssssssssssssss

The total jet luminosity can be inferred from observations using Eq.~(\ref{egcn}): 
\begin{equation}
L_{\rm j}=8\times 10^{44}\,L_{\gamma,41}^{5/8} M_{\rm BH,9}^{2/3}\theta_{-1}\,
\eta_{-1}^{-5/8}t_{\rm v,5}^{-1/8}M_{\rm RG}^{-1/6}\,\mbox{erg s}^{-1}.
\label{etj}
\end{equation}
This formula weakly depends on the observables, being almost insensitive to $M_{\rm RG}$, on the other hand hard to estimate.
This provides therefore a quite robust estimate of the jet luminosity with $\eta$ as most unknown parameter. Actually, if
$L_{\rm j}$ were known, then $\eta$ could be also estimated. 

For the most powerful jets, $L_\gamma$ would be limited by the jet size becoming $L_\gamma=\chi\eta L_{\rm j}$. 
Taking for instance $L_{\rm j}\sim 10^{47}$~erg~s$^{-1}$, $L_\gamma$ could
be as high as  $\approx 2\times 10^{45}\,\eta_{-1}$~erg~s$^{-1}$.  An improvement of a factor of several in the VHE
sensitivity (e.g. through the forthcoming the Cherenkov Telescope Array -CTA-) would test our gamma-ray predictions for the
whole RG-jet interaction process, including the early cloud expansion phase, allowing for a detailed study of the
involved (magneto)hydrodynamics, particle acceleration, and radiation.

We remark that, if a detectable gamma-ray flare with a duration of few days were to be produced in M87, in particular through $pp$
interactions, the cloud should have a mass of $\sim 10^{28}$~g. Such a massive cloud cannot acquire a large speed in the jet
direction at the times when $pp$ collisions are an efficient gamma-ray emitting mechanism, and therefore the emission will
not suffer significant Doppler boosting. In the case of a lighter cloud, large Lorentz factors can be achieved, but then $pp$
interactions will be inefficient producing gamma-rays, the probability to detect a flare lower due to beaming, and the
duration of the event shorter than observed because of faster expansion and beaming. 
%{  We note that for a blown cloud mass
%$>10^{28}$~g, the jet should be wider and the duration event would become longer.} 

{  Coming back to the question of cloud mass, we note that to extract a cloud with a mass $>10^{28}$ g, a more powerful 
jet than in M87, for similar jet-RG interaction conditions, would be required (see Eq.(\ref{mb})). }

An important question is whether there are enough RGs in M87 at the relevant jet scales. The model presented here would
require few interactions per year to explain the observations in M87. Since the typical duration of the RG-jet interaction is
of about 3--4~days, the RG filling factor should be $\Upsilon\sim 4/365\approx 10^{-2}$. With a jet volume at the relevant
scales of $\sim \pi\theta^2 z_{\rm T}^3/3$, the density of RGs in the region should be $\sim \Upsilon/V\sim 2\times
10^6$~pc$^{-3}$ for M87. Unfortunately, no direct information is available on the density of stars in the vicinity of the
SMBH in M87. The stellar mass in a sphere with a radius of 80~pc is estimated in $2\times 10^8 M_{\odot}$ 
\citep[e.g.][]{gt09}, and these observational data should be extrapolated four orders of magnitude down to $\sim 0.01$~pc.
Thus, depending on the assumed extrapolation law, the number of RGs in the vicinity of the SMBH may or may not be enough. It
is worth noting that a dense stellar cluster near the SMBH could be behind the broad-line region in AGN as produced by the
blown-up atmosphere of red dwarfs, which would imply the presence of numerous RGs in the center of AGN \citep{pens88}. In
addition, studies of the possible stellar density profiles in the vicinity of the SMBH in AGN \citep{bkck82,mcd91} show that
densities like the required one ($\sim  2\times 10^6$~pc$^{-3}$) could be achieved. The observation of VHE flares could be
already an indication that enough RGs are present near the SMBH in M87.

Interestingly, RG/jet interactions are expected to be transient phenomena. At higher jet heights, although many RGs could be
simultaneously present in the jet rendering rather continuous emission, the much more diluted jet would not remove a
significant amount of material from the star and the effective cross section of the interaction would be just $R_{\rm RG}$,
yielding a low energy budget for such a multiple interaction events.

The scenario presented here, adopted to explain the day VHE flares observed from M87, could also be relevant in other
non-blazar AGN. For blazar sources the beamed emission would overcome the RG-jet interaction, expected to be weakly beamed
due to moderate $v_{\rm z}$-values. For instance, the closest AGN, the radio galaxy Cen~A, at $\sim 3.8$~Mpc distance \citep{r04}, could
also show detectable flare like emission. At present, persistent faint VHE emission has been detected \citep{ah09} with
$L_{\gamma}=2.6\times 10^{39}\,\mbox{erg~s}^{-1}$. Accounting for the black hole mass of this AGN, $M_{\rm BH}=5.5\times
10^7\,M_{\odot}$, taking the observed VHE luminosity as a reference, and assuming $t_{\rm v}\sim 1$~day, one derives
implementing Eq.~\ref{etj} a jet luminosity $L_{\rm j}=1.2\times 10^{42}\,\mbox{erg~s}^{-1}$, a rather modest value.
Therefore, it cannot be excluded that RG-jet interactions may contribute to the VHE radiation detected from Cen~A, or that
transient activity due to RG-jet interactions may be observed from this source. Another case, the radio galaxy NGC1275, at a 
distance of 73~Mpc \citep{hwb09}, shows variable behaviour in GeV \citep{fer09}. The GeV luminosity is of about $2\times
10^{43}$~erg~s$^{-1}$. Using Eq.~(\ref{etj}), we can estimate the power of the jet as $5\times 10^{44}$~erg~s$^{-1}$ adopting
an $M_{BH}=10^8\,M_\odot$. In the case of NGC~1275 the shocked cloud would be optically thick at the luminosity peak,
implying significant attenuation of the TeV emission through photon-photon absorption with a cutoff around 50~GeV.

At farther distances, the strong jet luminosity dependence $L_{\gamma}\propto L_{\rm j}^{1.6}$ implies that FR~II sources
with say $L_{\rm j}\sim 10^{46}$~erg~s$^{-1}$ may be still detectable up to distances of $\sim 0.5$~Gpc (internal absorption
should be included in these cases; see \citealt{akc08}). Also, the luminosity in the range 0.1--100~GeV would also be
significant unless there is a strong low-energy cutoff in the proton spectrum. Therefore, Fermi may detect day-long GeV
flares originated due to RG-jet interactions from FR~II galaxies up to distances of few 100~Mpc. Summarizing, GeV and TeV
instrumentation can potentially detect a number of RG-jet interactions per year taking place in nearby FR~II and very nearby
FR~I galaxies, with the most powerful events being detectable up to 1~Gpc.

%%%%%%%%%%%%%%%%%%%%%%%%%%%%%%%%%%%%%%%%%%%%%%%%%%%%%%%%%%%%%%%%%
\section*{Acknowledgments}
V.B-R. wants to thank A. T. Araudo and G. E. Romero for fruitful discussions.
V.B-R. acknowledges support by the Ministerio de
Educaci\'on y Ciencia (Spain) under grant AYA 2007-68034-C03-01, FEDER funds. V.B-R. thanks Max Planck Institut fuer
Kernphysik for its kind hospitality and support. 
%%%%%%%%%%%%%%%%%%%%%%%%%%%%%%%%%%%%%%%%%%%%%%%%%%%%%%%%%%%%%%%%%

%\bibliographystyle{apj} % style aa.bst

%\bibliography{star_in_jet}

\end{document}